\documentclass[compsoc, conference, letterpaper, 10pt, times]{IEEEtran}

\newif\ifnamed\namedtrue

\newcommand{\myauthor}{Mario Werner, Thomas Unterluggauer, David Schaffenrath, and Stefan Mangard}
\newcommand{\myinstitute}{Graz University of Technology} 
\newcommand{\myemail}{\{firstname.lastname\}@iaik.tugraz.at}

\newcommand{\mytitle}{Sponge-Based Control-Flow Protection for IoT Devices}
\newcommand{\mykeywords}{control-flow protection, fault attacks, countermeasures, authenticated encryption, sponges}

\usepackage[utf8]{inputenc}
\usepackage[pdftex]{graphicx}
\graphicspath{{figures/}} 
\DeclareGraphicsExtensions{.pdf,.jpeg,.png}

\usepackage{url}
\usepackage{amsmath,amssymb}
\usepackage{csquotes} 
\usepackage{tikz} %
\usepackage{pgfplots}
\usetikzlibrary{calc,arrows,pgfplots.statistics} %
\usepackage{todonotes}  
\usepackage{booktabs}
\usepackage{balance}
\usepackage{gensymb}
\usepackage[caption=false,font=footnotesize]{subfig}
\usepackage{microtype}
\usepackage{footmisc}
\usepackage{cite}
\usepackage{wrapfig}           
\usepackage[export]{adjustbox} 

\let\oldtodo\todo
\renewcommand{\todo}[1]{\oldtodo[inline]{#1}}

\newcommand*\xor{\oplus}

\usepackage[style=long,nonumberlist,hyperfirst=false]{glossaries}

\usepackage{color,soul}

\usepackage[
unicode=true,
pdftex,
backref,
pagebackref=false,
bookmarks=true,
bookmarksopen=true,
pdfpagemode=UseOutlines,
plainpages=false,
colorlinks=false,
bookmarksdepth=3,
pdftitle={\mytitle},
pdfkeywords={\mykeywords}
]{hyperref}

\newacronym{ad}{AD}{Associated Data}
\newacronym{ae}{AE}{Authenticated Encryption}
\newacronym{aee}{AEE}{Authentic Encrypted Execution}
\newacronym{aeelight}{AEE-Light}{Authentic Encrypted Execution Light}
\newacronym{aslr}{ASLR}{Address-Space Layout Randomization}
\newacronym{bb}{BB}{Basic Block}
\newacronym{cfg}{CFG}{Control-Flow Graph}
\newacronym{cfi}{CFI}{Control-Flow Integrity}
\newacronym{csm}{CSM}{Continuous Signature Monitoring}
\newacronym{dep}{DEP}{Data Execution Prevention}
\newacronym{ddos}{DDoS}{Distributed Denial-of-Service}
\newacronym{dos}{DoS}{Denial-of-Service}
\newacronym{ie}{IE}{Infective Execution}
\newacronym{iot}{IoT}{Internet-of-Things}
\newacronym{ip}{IP}{Intellectual Property}
\newacronym{isa}{ISA}{Instruction-Set Architecture}
\newacronym{isr}{ISR}{Instruction-Set Randomization}
\newacronym{jop}{JOP}{Jump-Oriented Programming}
\newacronym{psa}{PSA}{Path-Signature Analysis}
\newacronym{ram}{RAM}{Random Access Memory}
\newacronym{rop}{ROP}{Return-Oriented Programming}
\newacronym{scfp}{SCFP}{Sponge-based Control-Flow Protection}
\newacronym{tee}{TEE}{Trusted Execution Environment}
\newacronym{tmto}{TMTO}{Time-Memory Trade-Off}
\newacronym{soc}{SoC}{System on Chip}
\newacronym{asic}{ASIC}{Application Specific Integrated Circuit}

\begin{document}

\title{\mytitle}

\ifnamed
  \hypersetup{pdfauthor={\myauthor}}
  \author{\IEEEauthorblockN{\myauthor}
  \IEEEauthorblockA{\myinstitute\\Email: \myemail}}
\else
  \hypersetup{pdfauthor={}}
  \author{\IEEEauthorblockN{}
  \IEEEauthorblockA\\ \\ \\\\}
\fi

\maketitle

\begin{abstract}
Embedded devices in the Internet of Things (IoT) face a wide variety of security
challenges. For example, software attackers perform code injection and code-reuse
attacks on their remote interfaces, and physical access to IoT devices allows to
tamper with code in memory, steal confidential Intellectual Property (IP), or
mount fault attacks to manipulate a CPU's control flow.

In this work, we present Sponge-based Control Flow Protection (SCFP). SCFP is a
stateful, sponge-based scheme to ensure the
confidentiality of software IP and its authentic execution on IoT devices. At
compile time, SCFP encrypts and authenticates software with instruction-level
granularity. During execution, an SCFP hardware extension between the CPU's
fetch and decode stage continuously decrypts and authenticates instructions.
Sponge-based authenticated encryption in SCFP yields fine-grained control-flow
integrity and thus prevents code-reuse, code-injection, and fault attacks on the
code and the control flow. In addition, SCFP withstands any modification of
software in memory.
For evaluation, we extended a \mbox{RISC-V} core with SCFP and fabricated a real
System on Chip (SoC). The average overhead in code size and execution time of
SCFP on this design is 19.8\,\% and 9.1\,\%, respectively, and thus meets the
requirements of embedded IoT devices.

\end{abstract}


\IEEEpeerreviewmaketitle

\section{Introduction}

\gls{iot} devices serve a variety of purposes, ranging from consumer products in
smart home environments, over sensor nodes in modern cars, to control units
in critical infrastructures. Typically, these embedded \gls{iot} devices
feature simple hard- and software architectures with only little consideration
of security to stay lightweight. However, the rapidly growing number of
\gls{iot} devices makes them an interesting target
for attackers. In particular, the security of pervasive \gls{iot} devices can
have direct impact on security and safety in the real world. For example,
the worm Stuxnet~\cite{falliere2011w32} spread across programmable logic
controllers in Iranian infrastructure in 2010, and the malware
Industroyer~\cite{IndustroyerWhitePaper2017} caused a black out of the
Ukrainian power grid in 2015. In addition, extensive \gls{ddos} attacks on
infrastructure providers by hijacking a large set of \gls{iot} devices, as with
the Mirai malware~\cite{MiraiBotnetSource2012}, are a significant threat.

There are numerous security challenges with \gls{iot} devices. The prevalent
Internet connection of \gls{iot} devices gives rise to remote attacks on their
exposed interfaces. In particular, attackers can try to find and exploit
vulnerabilities in these interfaces to take control over \gls{iot} devices
via code injection or code-reuse attacks, like
\gls{rop}~\cite{shacham_geometry_2007} and
\gls{jop}~\cite{bletsch_jumporiented_2011}. Further, many \gls{iot} devices
are run in hostile environments, where attackers have physical access to one or
many \gls{iot} devices. These physical attackers can read and tamper with
code in memory to perform code analysis and inject malicious
code. Access to code amplifies the risk of widespread code
injection and reuse attacks, and is a critical concern for software \gls{ip}
vendors. However, physical access also allows attackers to perform fault attacks
on the processor chip itself, by using, e.g.,~clock or power
glitches~\cite{DBLP:journals/iacr/Bar-ElCNTW04,korak_effects_2014,free60},
in order to manipulate the execution of code on the device. For example, by
skipping instructions using
power glitches, attackers can get control over one particular \gls{iot} device
to fake, e.g.,~sensitive sensor data.

To prevent code injection and code-reuse attacks, different countermeasures have
been proposed, such as \gls{dep}, return stack
protection~\cite{DBLP:conf/uss/Cowan98,Francillon:2009:DES:1655077.1655083,
DBLP:conf/osdi/KuznetsovSPCSS14},
\gls{aslr}~\cite{DBLP:conf/ccs/ShachamPPGMB04}, software
diversification~\cite{team2003pax} and \gls{cfi}~\cite{abadi_controlflow_2009}.
To protect the authenticity and confidentiality of \gls{ip}, encryption and
authentication of software binaries and \gls{ram} can be used. To counteract
physical fault attacks on the control flow of the processor,
\gls{cfi}~\cite{DBLP:conf/cardis/WernerWM15} is a feasible countermeasure as
well.

However, current embedded devices hardly implement any of these countermeasures.
Moreover, existing countermeasures work well for their original purpose in
isolation, but for each of them, some of the attacks on \gls{iot} devices remain
feasible due to the vast amount of different attack vectors. While a simple
combination of existing countermeasures can inhere overheads that are
impractical for lightweight embedded devices, the security analysis of
combinations of countermeasures can also become highly complex. Recently,
SOFIA~\cite{DBLP:conf/date/ClercqKC0MBPSV16,DBLP:journals/compsec/ClercqGUMV17}
was presented as the first approach
to counteract a combination of these attacks. By encrypting, authenticating and
chaining blocks of instructions using a stream cipher and MAC, SOFIA yields
\gls{cfi} as well as confidentiality and authenticity of software. However, one
drawback of SOFIA is its checking mechanism. In particular, the dedicated MAC
verification is a single point of failure that can potentially be exploited
using physical fault attacks on the error signal within the hardware. Furthermore,
the introduced code size and runtime overheads are potentially too high for certain
applications.

\subsubsection*{Contribution}
As an alternative approach to SOFIA and to overcome existing limitations, this
work presents \gls{scfp}.
\gls{scfp} is a novel, stateful scheme to protect the confidentiality of
software \gls{ip} and the authenticity of its execution in \gls{iot} devices. In
particular, \gls{scfp} encrypts and authenticates software binaries with
instruction-level granularity by using cryptographic sponges.
\gls{scfp} is designed as a hardware extension that continuously decrypts and
authenticates instructions in hardware at the latest possible point before the
processor's decode stage.

The use of sponge-based authenticated encryption in \gls{scfp} yields
fine-grained control-flow integrity and thus prevents code-reuse attacks. By
keeping the software encrypted throughout all memory, \gls{scfp} completely
thwarts code-injection attacks from within software, and effectively protects
the \gls{ip} of software vendors. By decrypting instructions right before the
decode stage of the processor, \gls{scfp} resists tampering with code in memory,
physical attacks on memory like
rowhammer~\cite{gruss_rowhammerjs_2015,kim_flipping_2014}, and fault attacks
that manipulate control flow or software code. \gls{scfp} supports 
interrupt handling
and is thus compatible with operating systems. Compared to existing work,
\gls{scfp} has lower memory and runtime overhead and offers strong fault
resistance. In
particular, any globally induced physical fault on the processor chip destroys
the internal \gls{scfp} state with high probability and leads to the execution
of random instructions, whereas
state-of-the-art \gls{cfi} schemes use a single verification step that can
be by-passed using controlled faults. Random code execution is a secure
processor state, because it is hard to control and exploit for an attacker, and
has a low probability of being meaningful. Furthermore, timely detection of
random execution is also supported.

\gls{scfp} is a highly flexible tool. We hence present two suitable sponge
constructions as well as three different
\gls{scfp} instances for different applications. First, \gls{aee}
provides all security features at cryptographic levels of security, i.e.,~above
80\,bits. Second, \gls{aeelight} reduces memory overhead in trade for reduced
software authenticity by using keyed permutations. Third, \gls{ie} is a very
lightweight \gls{cfi} scheme to solely protect against code-reuse and physical
fault attacks on the control flow.

\gls{scfp} is both practical and lightweight. For demonstration, we integrated
\gls{aeelight} with a \mbox{RISC-V} core and evaluated a set of benchmarks on
this processor by executing them both unprotected and encrypted with
\gls{aeelight}. It shows that the average overheads in code size and execution
time of our \gls{aeelight} instance are 19.8\,\% and 9.1\,\%, respectively, and thus
practical for many \gls{iot} scenarios.

\subsubsection*{Outline}

This paper is organized as follows. \autoref{sec:AEE} describes the concept of
\gls{scfp} and the application of authenticated encryption to the instruction
stream. \autoref{sec:mode} gives two sponge modes suitable for
\gls{scfp} and \autoref{sec:instantiation} presents different \gls{scfp}
instances and their security properties. \autoref{sec:evaluation} gives
evaluation results and \autoref{sec:related_work} provides a comparison with
related work. Finally, \autoref{sec:conclusion} concludes this work.

\section{Overall Concept}\label{sec:AEE}  

\glsfirst{scfp} is a novel security concept for \gls{iot} devices that is based
on authenticated encryption from cryptographic sponges. In this section, we
introduce the threat model we assume for \gls{iot} devices and present the
architecture of \gls{scfp}. In particular, we describe how sponge-based
authenticated encryption is applied to an instruction stream and discuss the
adaptions required to support arbitrary code execution including control-flow
transfers and interrupts.

\subsection{Threat Model and Assumptions}
This work considers \gls{iot} devices which are threatened by both software and
physical attacks. In terms of software vulnerabilities, we assume a remote
attacker who has arbitrary read and write access to the memory due to bugs in
the software. Correspondingly, active physical attackers are assumed to have
direct access to the device. This direct access can be used to dump and
manipulate external memory, to probe and force signals on the PCB (e.g.,~bus
signals between chips), or to inject global faults into the system (e.g.,~clock
glitches). On the other hand, micro probing and similar invasive techniques are
considered out of scope in this work. Similarly, side-channel leakage of hard-
and software implementations is not considered in this work.

Presumed targets for adversaries in this domain are to extract secret \gls{ip}
(e.g.,~firmware code), to bypass security checks (e.g.,~by skipping one or more
instructions), or to achieve arbitrary code execution via code reuse or
injection. In other words, adversaries try to compromise the confidentiality
and/or authenticity of the code, either at rest or at runtime. Note however,
that \gls{dos} as well as data-driven attacks are out of scope given that
neither can be solved via a \gls{cfi} scheme.

This work assumes that \gls{scfp} is deployed as the only countermeasure
to the mentioned threats. Hence, if guarantees that exceed the capabilities
of precisely enforced \gls{cfi} (e.g.,~resistance against control-flow
bending~\cite{DBLP:conf/uss/CarliniBPWG15}) are required, additional attack
mitigation techniques (e.g.,~safe stack~\cite{DBLP:conf/osdi/KuznetsovSPCSS14})
have to be utilized. Further, note that the hardware component of \gls{scfp} is
implemented in such a way that there is no interface to access plaintext
instructions, the sponge state or internal \gls{scfp} signals. All this
information is inaccessible in software.

\subsection{Architecture}

\begin{figure}[t]
  \centering
  \includegraphics[width=1.0\columnwidth,page=4]{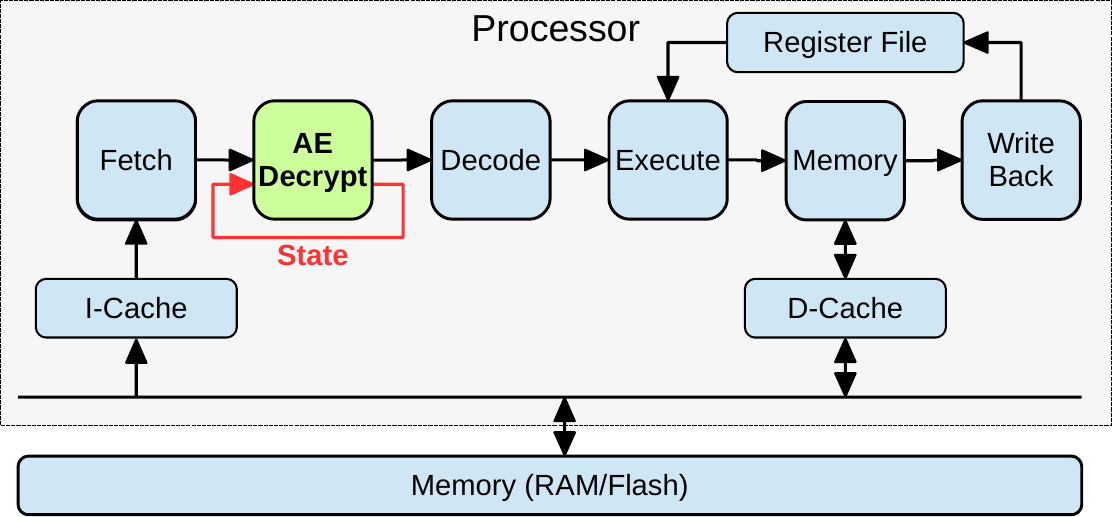}
  \caption{High-level system architecture of a classic RISC processor which has
  been extended for \gls{scfp} with a sponge-based \gls{ae} decryption stage.}
  \label{fig:arch_system}
\end{figure}

The idea behind \gls{scfp} is to encrypt programs at compile time using a
sponge-based \gls{ae} cipher. Decryption is then performed within the CPU,
instruction by instruction, just in time for execution. At its heart, the
sponge-based \gls{ae} cipher uses an internal state~$z$, which provides the
foundation for the \gls{cfi} protection in \gls{scfp}. This state accumulates
information about all the processed instruction ciphertexts, which enforces
that correct
decryption is only possible iff all previous instructions have
also been genuine. Conceptually, with every processed instruction
ciphertext~$C$, the plaintext instruction~$P$ as well as a new internal
state~$z'$ are derived from the current state~$z$ using a permutation~$f$
following $(P|z') = f(C|z)$ ($|$ denotes concatenation). As a result, the
correctness of the plaintext instructions that get executed by the CPU does not
only depend on the fetched input (i.e.,~ciphertext), but also on the history
which has been accumulated within the internal state of the cryptographic
primitive.

If either the state (e.g.,~through a \gls{cfi} violation or clock glitch) or the
ciphertext (e.g.,~through manipulation in memory) is erroneous, correct
decryption is not possible anymore and pseudo-random instructions are produced
as plaintext. We consider the respective execution of random instructions a
secure processor state for two reasons. First, the probability of random code
which is generated by \gls{scfp} to be meaningful is extremely low, especially
when attack gadgets of multiple instructions are required.
Second, attackers neither have control over the random instructions being
executed, nor can attackers observe what the plaintext
instructions are during random execution. This effectively hampers any attacker
attempts to execute harmful code. Besides, we will later show that \gls{scfp}
supports the detection of random code execution to add error handling as
desired.

From a processor architectural point of view, the ideal location within the
processor pipeline for performing the decryption is between instruction
fetching and decoding, as shown in \figurename~\ref{fig:arch_system}. The
instructions are transferred from the fetch to the decode stage exactly in the
execution sequence, which also matches the desired decryption sequence of
\gls{scfp}. As the decode stage is the first to need plaintext instructions,
performing decryption right in front of the decoder is in fact also the latest
possible point for inserting \gls{scfp} and effectively minimizes the number of
components with plaintext code access to the decode stage itself. All
the other components, like peripherals, main memory, various caches, memory
buses, and even the fetch unit, operate on encrypted code only.

\begin{figure}[t]
  \centering
  \includegraphics[width=1.0\columnwidth,page=3]{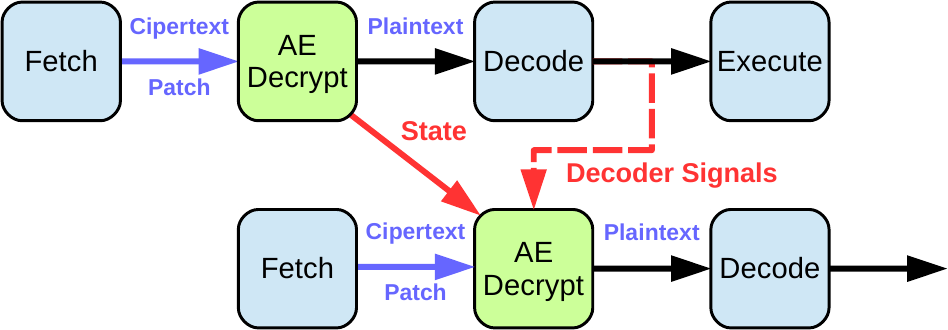}
  \caption{Data dependencies between two consecutive instructions within a
  processor pipeline when \gls{scfp} is implemented. The decoder signals can
  optionally be fed back.}
  \label{fig:arch_pipeline_dependendies}
\end{figure}

\figurename~\ref{fig:arch_pipeline_dependendies} depicts the instruction-data
dependencies between the different pipeline stages for the processor from
\figurename~\ref{fig:arch_system}. A traditional scalar processor with a
pipelined architecture only has dependencies between the different stages
(visualised horizontally) but not across multiple instructions (visualised
vertically). The processor basically decodes each instruction completely
isolated from other instructions. Dependencies between instructions are solely a
result of data dependencies in the program (e.g.,~via the register file) which
can lead to pipeline hazards and stalls. Extending the pipeline with an \gls{ae}
decryption unit breaks this isolation between instructions and introduces an
additional dependency via the cipher state.

For scalar processors, it is additionally possible to feed the data independent
decoder signals
of each executed instruction back into the cipher. Such feedback extends
authenticity protection up to the pipeline's execute stage and can, for example,
be used as a link to fault countermeasures in the ALU. Note however, that the
\gls{scfp} approach is not limited to scalar processors. Superscalar
microarchitectures can also be protected using \gls{scfp} with a coarser
granularity, e.g.,~decrypting multiple instructions instead of individual
instructions in one block.

\subsection{Authenticated Encryption and Control Flow}

Sponge-based authenticated encryption schemes use a single internal cipher state
for both encryption and authentication. This common state leads to the nice
property that the mapping between each encrypted and plain instruction depends
on the actual values of all previously processed instructions. Hence, to be able
to encrypt a program such that it can be executed on a processor that implements
\gls{scfp}, the exact sequence of executed instructions needs to be known at
compile time. However, exactly this property makes the combination of
authenticated encryption with control flow challenging.

More concretely, at compile time, the exact instruction sequence can only be
determined for a very limited number of programs. Basically, only programs that
have a completely data independent control flow (e.g.,~no data dependent
branches) can be trivially supported. Additionally, even genuine and intended
code reuse (e.g.,~loop bodies or functions) is not easily possible anymore. This
is due to the fact that after encryption, the ciphertext is fixed and correct
decryption of an instruction is only possible given the correct unique cipher
state (and thus execution history). Placing the sponge-based authenticated
encryption scheme into the processor pipeline therefore provides a solid
foundation for \gls{scfp} and thwarts code reuse by default.

The main idea to allow specific code reuse in \gls{scfp} and to make \gls{scfp}
applicable to general programs is to deliberately introduce collisions into the
internal state of the cryptographic primitive. These state collisions are
conceptually a white listing of permitted control flow transfers and have to be
introduced exactly at the required positions in a program. Note that these
deliberately introduced collisions do not weaken the security of the
cryptographic primitive.

The simplest and most efficient way to generate the required state collisions is
to inject additional metadata as correction terms into the cipher state at
certain points during the execution of the program. We denote this process of
deliberately adjusting the \gls{ae} state as \emph{patching} and the involved
constants as \emph{patch values}. Via patching, we effectively cancel out
divergences in the cipher state which originate from taking different valid
paths through the \gls{cfg}. As the result, correct decryption of a program
under \gls{scfp} is only possible as long as the execution adheres to the
statically determined \gls{cfg}.

It has to be noted that patching must be implemented as a differential update of
the \gls{ae} state. Otherwise, if patching was implemented by simple
replacement, patching would destroy all the history which had been accumulated
into the state. Besides, the patching process must be able to modify the full
sponge state in order to create arbitrary collisions.

\subsection{Patch Handling, Placement and Calculation}

The patch values in \gls{scfp} are conceptually very similar to the justifying
signatures in the soft error and fault attack countermeasures based on
\gls{csm}~\cite{DBLP:journals/tcad/WilkenS90,DBLP:conf/cardis/WernerWM15}.
Therefore, also similar implementation techniques can be used to find suitable
patch locations as well as to determine the concrete values of the
patch constants.

\begin{figure}[t]
  \centering
  \includegraphics[width=0.35\columnwidth,page=5]{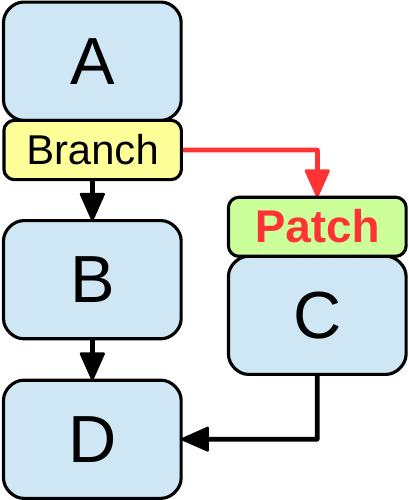}
  \caption{\label{fig:StatePatching_if}Simple example of patching the \gls{cfg}
  of an if-then-else construct in \gls{scfp}.}
\end{figure}

More concretely, the task of the patch values in \gls{scfp} is to introduce
cipher state collisions at the merge points in the \gls{cfg} of the program.
Hence, all differences which originate from traversing the statically determined
\gls{cfg} along runtime data dependent paths have to be compensated. An example
for patching a simple if-then-else construct is shown in
\figurename~\ref{fig:StatePatching_if}. There, a patch value is injected into
the cipher state before the execution of \gls{bb} C (i.e.,~on the red \gls{cfg}
edge) such that the state at the beginning of \gls{bb} D is the same, regardless
of whether the blocks A and B, or the blocks A and C (incl. the patch) have been
executed.

The exact way how such a patch value is encoded into the program and how patches
are processed during runtime strongly depends on the concrete implementation of
\gls{scfp} and is highly \gls{isa} specific. However, an intuitive way to
implement and think about cipher state patching is to consider the patch values
as part of specialized control-flow instructions. Similar to immediate operands
in standard instructions, the patch values are part of the instruction encoding
and get fetched like regular code by the processor during execution.

From the toolchain perspective, implementing \gls{scfp} consists of two steps.
In the first step, during compilation, patches have to be inserted at the
correct positions into the program by emitting suitable instructions with patch
support. In the second step, at link time or in a post processing phase, the
program binary has to be encrypted and the correct patch values have to be
inserted into the binary (i.e.,~similar to relocations).

For a program which comprises only branches and direct calls, a functional
solution for patch placement during compilation can be obtained by looking at
the undirected \gls{cfg} of the full program. Every cycle in this graph has to
be broken by introducing a patch for the cipher state. Therefore, the minimum
number of patches and possible positions can be obtained by comparing the
\gls{cfg} with its spanning tree. Taking the function call graph into account,
this approach is also applicable to indirect and recursive function calls.
Unfortunately, comparably expensive whole program analysis has to be performed
to acquire the mentioned graphs.

Nevertheless, also compilation in multiple translation units can be supported
with \gls{scfp} when a well-defined patching convention is established around
function calls. Similar to a regular calling convention, having a
patching convention allows to correctly place patches in every function of the
program in isolation. Within each function, it is then typically sufficient to
always patch when a branch is taken as shown in
\figurename~\ref{fig:StatePatching_if}. Additionally, to cope with recursion, it
has to be ensured that at least one patch is performed before the recursion is
entered. Note however that the simplicity of the patching convention, compared
to the graph based approach, comes at the cost of an increased number of patch
values.

To illustrate the concept, in the following, an exemplary patching convention
for direct and indirect function calls is presented.

\subsubsection{Direct Calls}

\begin{figure}[t]
  \centering
  \includegraphics[width=0.69\columnwidth,page=6]{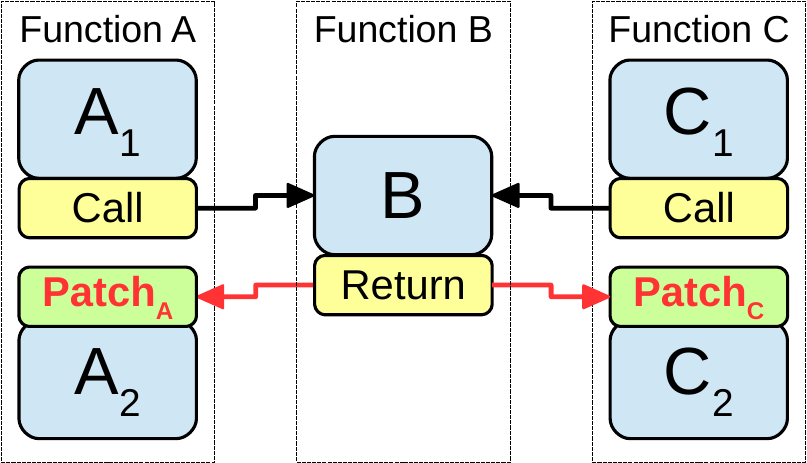}
  \caption{\label{fig:StatePatching_call}Example of a simple patching convention
  for direct function calls. Function B can be called from both, function A and
  C.}
\end{figure}

Every function which gets directly called from more than one call site within a
program necessarily requires patching. In particular, at least $n-1$ patches are
required when $n$ call sites exist. Interestingly, this situation is also
similar to the direct branch example in \figurename~\ref{fig:StatePatching_if}
where one patch is required since two paths in the \gls{cfg} lead to \gls{bb} D.
However, placing patches at every call site except one again requires access to
the full program during compilation. To relax this constraint, at the cost of
one additional patch per function, patching can simply be performed on every
call site as shown in \figurename~\ref{fig:StatePatching_call}. In this example,
$Patch_A$ has to be applied when the control flow returns from function B to
function A. Returning from function B to C uses $Patch_C$, respectively.

Note that, in most cases, having one patch per direct call is sufficient
regarding both functionality and security, because typical \glspl{isa} perform
direct calls relative to the program counter. In this case, the program counter
relative offset is part of the function call encoding and is different for each
call site. This implies a different, internal \gls{scfp} state for each call
site. As a result, besides the required state collisions at the call
and return edges of the direct function call, there are no other, undesired
collisions being introduced to the program.

In general, it does not matter whether the patch value is applied at the return
operation or the call operation, as long as it is done consistently and aligned
with the way branches are patched. Applying patches is therefore possible either
after branches and on returns (as shown in
\figurename~\ref{fig:StatePatching_if} and
\figurename~\ref{fig:StatePatching_call}), or before merge points of branches
and during calls. In fact, when looking at the \gls{cfg} of the whole program,
function returns are simply branches and function calls are merging points.

\subsubsection{Indirect Calls}

Similar to direct function calls, also indirect function calls require patching.
However, determining the exact function which gets called at runtime by an
indirect function call is not always possible at compile time. Moreover, often
also multiple different functions get called from the same indirect call site
during the runtime of a program (e.g.,~comparison callback of \texttt{qsort}).
Therefore, the best one can do with static \gls{cfi} such as \gls{scfp} is to
determine a, possibly over approximated, set of potential call targets and to
enforce that only calls to functions in this set are possible at runtime.

\begin{figure}[t]
  \centering
  \includegraphics[width=0.8\columnwidth,page=8]{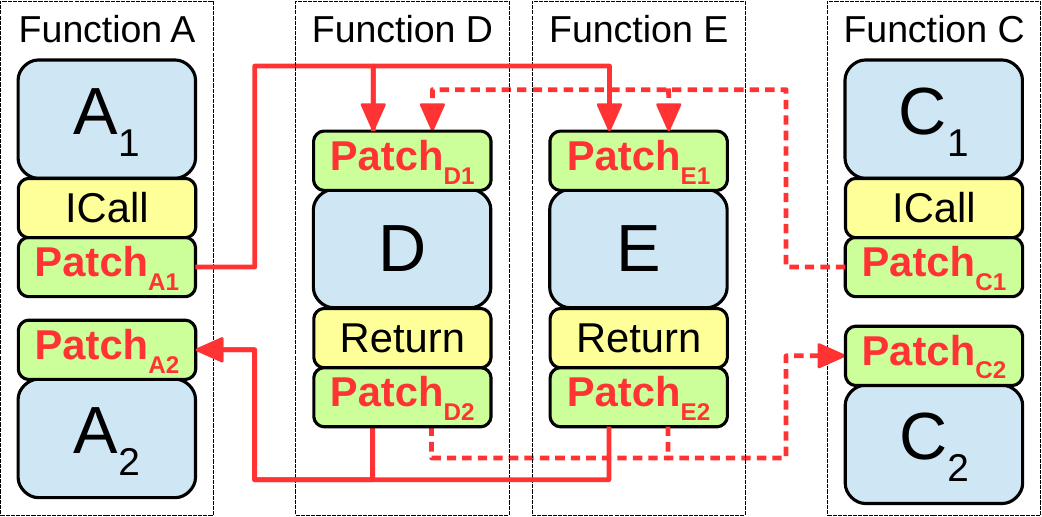}
  \caption{\label{fig:StatePatching_icall}Example of a simple patching
  convention for indirect function calls. Functions A and C can call both,
  functions D and E at runtime.}
\end{figure}

Our current approach to implement indirect function calls and returns with
\gls{scfp} is shown in \figurename~\ref{fig:StatePatching_icall}. In total, two
patch values have to be applied on every indirect control-flow transfer. The
idea of this scheme is to use the first patch (e.g.,~$Patch_{A1}$) to reach a
constant cryptographic intermediate state, which is then updated to the actual
entry state of the called function using the second patch value
(e.g.,~$Patch_{D1}$). The constant intermediate state can be freely chosen at
encryption time and permits to restrict indirect calls to targets which were
encrypted for the same intermediate state.

In summary, for the patching convention in
\figurename~\ref{fig:StatePatching_icall}, two patch values are required for
every indirect function call site as well as for every function which can be
called indirectly. At runtime, in total four patches get applied for every
indirect function call.

At the first glance, using four patch values for one indirect function call may
seem excessive given that two patches would already suffice to build a
functioning \gls{cfi} scheme. However, using less patch values necessarily
introduces undesired collisions into the \gls{scfp} state which weakens the
confidentiality and authenticity properties of the scheme.

\subsection{Initial State Derivation}

In sponge-based \gls{ae} ciphers with known permutation, the initial
state is comparable to the key in regular encryption schemes. It is
common to derive this initial state $z_I$ from a secret key $k$ and public nonce
$N$ by applying their permutation $f$ (e.g.,~$z_I = f(N|k)$). Conceptually, we
recommend using a similar approach for deriving the initial state in
\gls{scfp}. This ensures that, even when $k$ is a device-specific fixed master
key, every program for that device is still encrypted under a different initial
state. Optionally, additional information like, for example, the start address
or the program vendor can be used during the derivation of the initial state.

Note that binding initial states to the machine key also serves as software
diversification. Namely, in case successful exploits against \gls{scfp} should
be found in a program on a certain device, they cannot simply be transferred to
other devices executing the same program.

\subsection{Interrupt Handling}

Unlike regular function calls, which are performed at precisely defined points
during program execution, interrupts can occur at virtually any time. It is
therefore impossible to determine a unique differential update value for all
the states which permit to call an interrupt handler. We cope with this problem
in \gls{scfp} by treating interrupt handlers similar to the initial program
entry point. Therefore, we derive a new \gls{ae} cipher state to re-initialize
\gls{scfp} when entering the interrupt handler. On the other hand, the
\gls{scfp} state that is active before entering the interrupt handler is,
similar to the old program counter, saved in an internal processor register.
For the operating system, the \gls{scfp} state is therefore simply one
additional register which has to be saved and restored during context switches.
Note however that, to ensure that the confidentiality of the \gls{scfp}
state is maintained at all times, the old state value which is stored in the
processor register should be encrypted or similarly protected.

Implementing interrupt handling in this way effectively separates the protection
of interrupt handlers from the regular code. This means that interrupts can be
processed successfully even when a regular program executes pseudo random
instructions due to an attack. On handler entry, this separation is desirable
as it allows us to recover from errors in software as well as to perform
scheduling of programs via the operating system. On handler exit, on the other
hand, we want to propagate errors occurring during the execution of the
interrupt handler into the execution of the regular code.

We achieve this behavior by enforcing that the internal \gls{scfp} state has a
predefined secret value when returning from the interrupt handler. Similar to
the state derivation on interrupt entry, the secret handler exit state can, for
example, again be computed from the key, the nonce, and the address of the
interrupt handler. When returning to the regular code execution, the hardware
can then simply combine the current state $z$, the expected exit state $e$, and
the state from before the interrupt entry $z_{entry}$ from the register to
calculate the next state $z'$, e.g.,~$z' = z \xor e \xor z_{entry}$. By doing
so, the entry value is only restored ($z' = z_{entry}$) correctly as next state
if also the handler execution has been genuine ($z == e$).

\subsection{Fast Error Recovery}

As \gls{scfp} ensures security even without explicit fault checks, \gls{scfp}
eliminates the existence of a single point of failure. Namely, the
probability of random code execution in \gls{scfp} to be meaningful is extremely
low. While this is one major benefit of \gls{scfp}, it may still be desirable
to provide a timely way to perform error recovery after the processor started to
execute a random instruction sequence. Interestingly, the execution of pseudo
random instructions
in the error case already provides one way to permit error recovery since the
processor is able to identify invalid instructions. The concrete detection
probability follows a geometric distribution and can be computed when the
\gls{isa} of the processor is known. More concretely, given the probability
$p_{inv}$ for
a random instruction to be invalid, the expected detection latency $l$
is computed as $l = 1/p_{inv}$. However, considering that modern \glspl{isa} are
often quite dense, recovery latency can be comparably high.

Faster recovery can be achieved when additional redundancy bits are verified on
the execution of every single instruction. Sponges permit to implement this
additional integrity verification in an efficient and secure way by simply
checking the desired amount of state bits. No additional permutation calls, but
only a marginally bigger permutation is required. The strength, i.e.,~the
number of bits, for this verification can be freely chosen, but is typically
rather weak for a single instruction. However, the continuous nature of this
check compensates for this weakness quite fast. In general, the number of
asserted bits allows to trade off between the code size overhead and the
recovery latency.

\section{Sponge Constructions for \gls{scfp}}\label{sec:mode}

\gls{scfp} relies on a scalable and strong sponge-based authenticated
encryption cipher. This section introduces two eligible sponge-based
constructions and presents arguments for their security as well as guidelines
for parameter selection.

\subsection{Constructions} 

Cryptographic sponges have become quite popular since Keccak has been announced
as the winner of the \mbox{SHA-3} competition. However, sponges can also be
used to build other cryptographic primitives. The Keccak designers themselves,
for example, already proposed an \gls{ae} mode called
SpongeWrap~\cite{bertoni_duplexing_2011} early on and still pursue the idea with
Keyak~\cite{bertoni_keyak_2015} and Ketje~\cite{bertoni_ketje_2014} in the
ongoing CAESAR competition~\cite{bernstein_caesar_2016}. The other numerous
sponge-based submissions~\cite{andreeva_primates_2014, aumasson_norx_2014,
dobraunig_ascon_2015, gligoroski_cipher_2015, morawiecki_icepole_2014,
saarinen_stribobr2_2015} to the competition further support this research
direction.

Considering the success and general properties of sponges, the following
discusses two sponge-based constructions which have been adapted to support the
patching of \gls{scfp}. This approach allows us to profit from the substantial
amount of cryptanalysis performed on the various sponge constructions and the
underlying permutations. In general, we therefore recommend well-analyzed
permutations like \mbox{Keccak-$p$}. However, a more detailed discussion on
suitable instantiations of \gls{scfp}, including permutations, can be found in
\autoref{sec:instantiation}.

\subsubsection{SpongeWrap-like Decryption Mode}

\begin{figure}[t]
  \centering
  \includegraphics[width=1.\columnwidth,page=9]{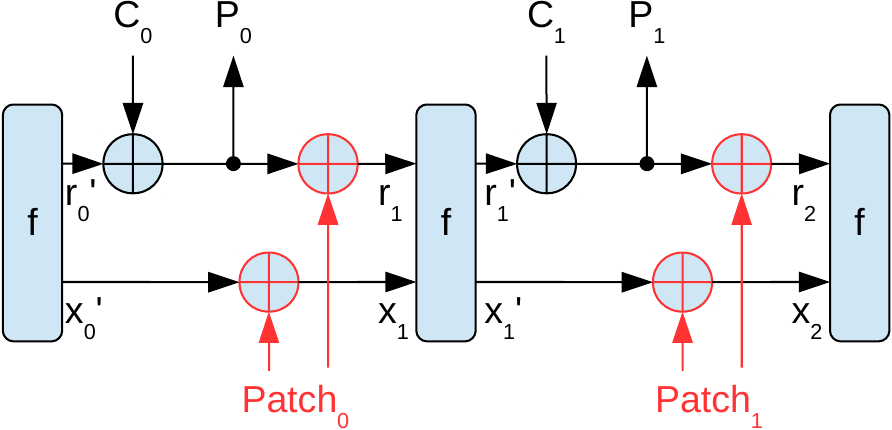}
  \caption{Decryption using a duplex construction similar to the one used in
  SpongeWrap.}
  \label{fig:sponge_decryption_duplex}
\end{figure}

The first construction, shown in \figurename~\ref{fig:sponge_decryption_duplex},
is based on the duplex construction, which has been introduced and proven to be
secure in~\cite{bertoni_duplexing_2011}. This duplex construction is used in
SpongeWrap for both encryption and decryption. When executing strictly
sequential code, where no patching is required ($Patch_i = 0$), \gls{ae} on the
instruction stream is identical to \gls{scfp}. However, for generic code
\gls{scfp} must also implement branching. Therefore, additional support
for the injection of patch values has to be added to the construction. Both the
rate and the capacity of sponge make up the previously described \gls{scfp}
state $z$ and must be modifiable by such a patch.

From the security point of view, these patch values can be considered as
\gls{ad}. \gls{ad} denotes data that is authenticated, but not encrypted. It
has been shown by Mennink et al.~\cite{mennink_security_2015} as well as Sasaki
and Yasuda~\cite{sasaki_how_2015} that it is secure to absorb \gls{ad} into the
capacity of a keyed sponge. Considering that the construction in
\figurename~\ref{fig:sponge_decryption_duplex} is a keyed full-state duplex
sponge construction, it is therefore secure to inject the patch values into the
capacity. Updating the rate with the patch is secure as well given that the rate
is under the control of an attacker via the ciphertext in any case.

The SpongeWrap-like construction has two neat features. First, its
implementation is comparably simple since encryption is identical to
decryption. Second, the construction provides great flexibility as it permits to
calculate and place patch values on arbitrary places in the \gls{cfg}. However,
there
are also some drawbacks which have to be considered. For example, an attacker
might be able to precisely control the first fault given that errors in the
ciphertext directly propagate into the plaintext ($\Delta C_i = \Delta P_i$).
In a known plaintext attack, this might even permit to inject one specific
instruction before the plaintexts of subsequent instructions are randomized.

Note also that, if the control-flow merges at instruction $i$ and patches are
not applied directly before the merge point in the control-flow graph
(i.e.,~$Patch_{i-1}=0$), all instructions directly preceding the merge point
($P_{i-1}$) have to be identical. This is due to the fact that, as soon as the
instruction at the merge point is fixed (i.e.,~$P_i$ and $C_i$), also the
plaintext of the predecessor $P_{i-1}$ is determined by the dependency over the
rate part of the sponge ($P_i = C_i \oplus f_r(P_{i-1} | x_{i-1}')$). However,
this link can be broken by performing patches solely at merge points of the
\gls{cfg} instead of placing them freely.

\subsubsection{APE-like Decryption Mode}

\begin{figure}[t]
  \centering
  \includegraphics[width=0.8\columnwidth,page=6]{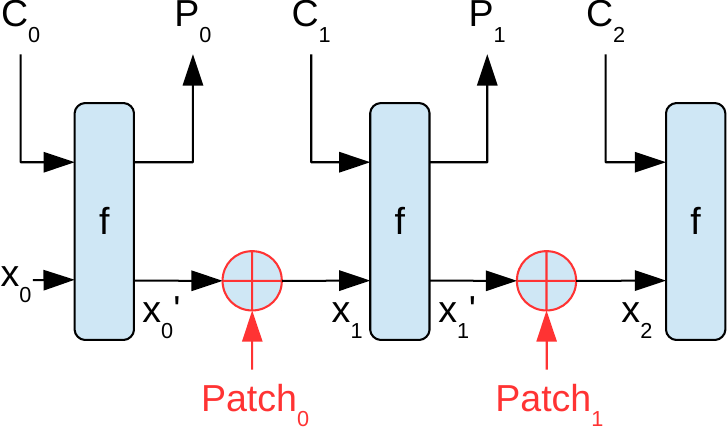}
  \caption{Decryption in an APE-like construction.}
  \label{fig:sponge_decryption_ape_light}
\end{figure}

The second construction is inspired by another \gls{ae} mode of operation called
APE~\cite{andreeva_ape_2014}. The layout of the APE construction itself is
similar to the duplex construction in SpongeWrap. However, APE is not inverse
free, i.e.,~the inverse permutation $f^{-1}$ is needed for encryption when the
permutation $f$ is used for decryption. Moreover, the indices of the cipher-
and plaintexts have been rearranged compared to SpongeWrap. Namely, in APE, the
plaintext $P_i$, corresponding to a ciphertext $C_i$, is calculated as $P_i =
C_{i+1} \oplus f_r(C_{i} | x_i)$.

The APE-inspired mode we propose in
\figurename~\ref{fig:sponge_decryption_ape_light} is calculated as $P_i =
f_r(C_{i} | x_i)$ and modifies APE in two ways. First, $P_i$'s dependency on
$C_{i+1}$ is removed. This solves the problem of the SpongeWrap-like mode where
attackers can inject one specific instruction if they know the original value.
Moreover, this modification makes our construction behave more like a block
cipher than a stream cipher. Second, patching capabilities for the capacity are
introduced to make the construction suitable for \gls{scfp}. Note that the
APE-like construction is superior to the SpongeWrap-like mode in this regard. It
only needs patching of the sponge's capacity which corresponds to the \gls{scfp} state $z$. On the other hand, the sponge rate is not chained any more.

The main drawback of the APE-like construction is that it is less flexible,
because the position of patches is fixed. Patches always have to be positioned
at branching points in the control-flow graph. This is a result of the
encryption that has to be performed in inverse direction to the decryption
(i.e.,~inverse to the execution sequence).

\subsection{Parameter selection} 

It has been shown that the duplex construction~\cite{bertoni_duplexing_2011} as
well as the APE construction~\cite{andreeva_ape_2014} are secure against generic
attacks which do not exploit properties of the underlying permutation. The
complexity of such attacks is lower bounded by $2^{x/2}$ and depends on the
capacity size $x$. To provide $s$-bit security, $x$ must thus be chosen as $x
\ge 2 \cdot s$.

The size of the sponge rate depends on the actual implementation. The majority
of the rate is needed for the decryption of the instructions. The instruction
size $i$ depends on the \gls{isa} and is typically 16 or 32 bits. However,
additional rate bits may be used for fast error recovery. To enable fast error
recovery of $n$ bits without leaking parts of the plaintext nor reduction of the
security, a rate of $r = i + n$ bits, and hence a permutation size of $b
= i + n + x$ bits is needed.

The proofs in~\cite{andreeva_security_2015,bertoni_security_2011} show that
also smaller capacity sizes can result in cryptographic security.
These results can be used to either reduce the permutation size while
maintaining the security level, or to increase the security of a fixed
permutation. However, a limit on the data complexity, which strongly depends on
the actual implementation, is required to benefit from these refined proofs. We
therefore refrain from proposing parameters based on the proofs
in~\cite{andreeva_security_2015,bertoni_security_2011} and leave the
exploitation to implementers knowing the respective system characteristics.

\section{Instantiations}
\label{sec:instantiation}

The flexibility of \gls{scfp} allows to tailor its protection level to the
needs of the respective application by choosing a suitable permutation
for the sponge-based \gls{ae} scheme.
In this section, we hence introduce three different instantiations of
\gls{scfp}.
First, \gls{aee} uses a large, unkeyed permutation to yield confidentiality and
authenticity of the program binary as well as \gls{cfi} to prevent fault, code
reuse, and code-injection attacks. Second, \gls{ie} uses a small, unkeyed
permutation to form a lightweight \gls{cfi} scheme to prevent code-reuse and
fault attacks only. Third, the use of a small, keyed
permutation in \gls{aeelight} yields small overhead, \gls{cfi} and \gls{ip}
protection in trade for weaker authenticity. We first discuss the properties of
unkeyed permutations used in \gls{aee} and \gls{ie}, and then proceed with keyed
permutations utilized in \gls{aeelight}.

\subsection{Unkeyed Permutations}

When instantiating \gls{scfp} with unkeyed permutations, the cryptographic
security properties of \gls{scfp} are solely determined by the size of the
sponge capacity $x$. Neglecting the proofs
in~\cite{andreeva_security_2015,bertoni_security_2011}, a security level of
$s$\,bits requires a sponge capacity of $2s$\,bits. However, these are generic
results without consideration of the actual application.

In particular, the cryptographic security level $s$ is mainly determined by
collisions in the cryptographic state. These can generically be exploited in
\gls{tmto} attacks with birthday bound complexity $2^{x/2}$ and eventually
allow state recovery and thus \gls{ip} theft or forgery. However, to perform
these \gls{tmto} attacks, the attacker must
also be able to observe the output of the sponge, which is not the case for
\gls{scfp}. Namely, the instructions decrypted by \gls{scfp} are internally
processed by the processor and never directly revealed to the attacker. As a
result, the
complexity for state recovery for \gls{scfp} is $2^x$ in practice.
In a similar way, the probability of arbitrary state collisions in a binary
encrypted and authenticated with \gls{scfp} is in general determined by the
birthday bound, i.e.,~$2^{-x/2}$. However, attackers do not have access to the
decrypted instructions and the internal state when using \gls{scfp}. Attackers
are thus unable to
observe and detect internal state collisions. Hence, meaningful exploitation
of internal state collisions
for \gls{scfp} is equivalent to state recovery and has complexity $2^x$ as well.

\textbf{Software Attack Complexity.}
These considerations have a significant impact on the actual attack
complexities for code injection and code-reuse attacks when \gls{scfp} is in
place, i.e.,~the \gls{cfi} properties of \gls{scfp}. Namely, attackers
performing code injection or code-reuse attacks require precise control over
the executed instructions to succeed.
For example, attackers can modify a single instruction with success probability
$2^r$, but will neither be able to observe whether they hit the right
instruction, nor be able to modify the internal state such that all successive
instructions remain the same. This means that the attacker must adapt all
successive instructions too, because the processor will otherwise execute random
instructions. However, precise manipulation of $n$ instructions has even higher
complexity, namely $2^{n \cdot r}$. Alternatively, attackers can try to learn
the internal state to correctly encrypt and inject their own program. However,
this has complexity $2^x$. A different example are modified jump
targets in code-reuse attacks. As attackers manipulate addresses to jump to
well-defined instructions in the binary, the $x$-bit patch values must be
adapted accordingly. However, finding a correct patch value has complexity
$2^x$ too.
 
\textbf{Fault Attack Complexity.}
The \gls{cfi} properties of \gls{scfp} also increase the attack complexities
for fault injection attacks that manipulate control flow or instructions prior
to the decode stage. For example, simple instruction skips or repetition have a
success probability of $2^{-x}$.  The same probability applies to arbitrary
control-flow errors, e.g.,~caused by a randomly faulted program counter. On the
other hand, performing a specific control-flow transfer via faults is as hard
as forcing the $x$ capacity bits and the program counter (e.g.,~32~bits) to the
desired value. However, this is non-trivial since the sponge state is secret and
must be extracted or brute forced first. Furthermore, being able to control that
many bits precisely is quite hard in practice.

Instead of altering the control flow, fault attackers can also try to
manipulate code by injecting bit flips. For example, attackers can use clock or
power glitches to inject random bit flips into code. However, it takes roughly
$2^{r}$ tries to hit one specific instruction with random bit flips. Therefore,
another approach is to use a small and limited number of precise bit flips in
the fault attack instead. Yet, exploiting precise bit flips in the encoded value
is as hard as utilizing a differential characteristic of the permutation. Only
precise bit flips in the plain instruction can be exploited directly. However,
regardless of whether random or precise bit flips are injected, bit flips in
code modify the sponge state as well. This randomizes the sponge output of all
subsequent instructions and therefore prevents further exploitation. Moreover,
\gls{scfp} can also protect the plain instructions against fault
attacks as well by feeding the decoder signals back into the sponge state.

Depending on the concrete security level $s$ and choice of sponge parameters $r$
and $x$, we identify two different types of \gls{scfp} instances using unkeyed
permutations. First, \gls{aee} denotes instances with cryptographic security
levels, i.e.,~at least 80\,bits, that offer \gls{cfi} as well as confidentiality
and authenticity of software \gls{ip}. Second, \gls{ie} denotes instances below
cryptographic security levels to enforce \gls{cfi} only.

\subsubsection{\glsdesc{aee}}

\gls{aee} features cryptographic security levels for encrypting and
authenticating code. This automatically defeats adversaries which wiretap
the communication to the external memory chips without any need for further code
encryption and/or authentication. Moreover, software attacks are made harder
too. As other \gls{cfi} schemes, \gls{aee} hampers return- and jump-oriented
programming attacks. The strong encryption and authentication further mitigates
both code injection and code-disclosure attacks. \gls{aee} is therefore a
replacement for established software attack countermeasures like DEP/W\^{}X, CFI, and
R\^{}X. In addition, by enforcing \gls{cfi} \gls{aee} also prevents fault
attacks on the processor chip that aim at instruction or control-flow
manipulation.

From a cryptographic perspective, \gls{aee} requires a permutation size of at
least 192\,bits to yield 80-bit security for a 32-bit instruction
set. One suitable permutation to instantiate \gls{aee} hence is
\mbox{Keccak-$p$[200,12]}~\cite{bertoni_keyak_2015} with 200-bit state size
and 12 rounds as used in Keyak. The exceeding 8\,bits increase the capacity and
thus the security level to 84\,bits. However, as elaborated before, the
specifics of \gls{aee} result in a complexity of $2^{168}$ for state recovery,
control-flow hijacking, and fault attacks on control flow. Similarly, a single
instruction can be successfully manipulated from software or using fault
attacks with complexity $2^{32}$, but the internal 168-bit state will cause the
execution of random instructions afterwards.

\subsubsection{\glsdesc{ie}}

Contrary to \gls{aee}, \gls{ie} uses a small permutation and thus, from a
cryptographic point of view, cannot provide a strong level of security.
In particular, \gls{ie} behaves like a context-sensitive \gls{isr} rather than
authenticated encryption. \gls{ie} thus fails to
ensure confidentiality and authenticity of software \gls{ip}. However, the
parameterization of \gls{ie} forms a practical \gls{cfi} scheme that
considerably complicates code-reuse attacks as well as fault attacks on the
processor chip itself. Yet, the concrete instantiation of \gls{ie} is highly
application specific.

For a 32-bit \gls{isa}, \gls{ie} can, for example, be instantiated with 50-bit
state size and the \mbox{Keccak-$p$[50,12]}~\cite{bertoni_keyak_2015}
permutation (i.e.,~12 rounds as in Keyak). Using two bits for fast error
recovery gives a sponge rate $r=34$\,bits and a sponge capacity $x=16$\,bits,
which also corresponds to the size of the patch values. From a cryptographic
perspective, this \gls{ie} instance yields merely 8-bit security.
However, the probability for successful code injection and manipulation of
control flow still is $2^{-16}$.

The main drawback of \gls{ie} is that an attacker with access to the encrypted
binary can easily perform state recovery offline, in our example with complexity
$2^{16}$. State recovery eventually breaks the \gls{cfi} property of \gls{ie}
for software attackers. Namely, a software attacker knowing the secret, internal
\gls{ie} state can compute correct ciphertexts and patch values, and inject
these into the code from within software when performing code injection or reuse
attacks. However, the complexity of physical fault injection on the processor
chip itself is still high enough for the parameterization of \gls{ie}.
Nevertheless, to ensure \gls{cfi} for software attackers as well, access to the
encrypted binary must be limited. While this restricts the attacker compared to
the original threat model in \autoref{sec:AEE}, access control can easily be
enforced using two different mechanisms: (1) by using execute-only memory,
software attackers lose online access to the encrypted binary, and (2) by storing
the binary in on-chip memory, attackers with physical access cannot read the
encrypted binary any more. As a result, \gls{ie} is particularly interesting for
tiny \gls{iot} devices without external memory and for smart cards.
Note however that state recovery, code analysis, and wide-spread
deployment of attacks can easily be mitigated by using a different seed for
\gls{ie} on every device as this causes the internal states, patch values,
ciphertexts, and positions of state collisions to change. Moreover, note that
the probabilities for manipulating control flow stated above are enough to
enforce \gls{cfi} and are indeed in the range of entropy estimations of other
techniques to prevent code-reuse attacks, e.g.,~software
diversification~\cite{DBLP:conf/sp/ClementsASSKBP17}.

\begin{table*}[t]
 \caption{Examples of \gls{scfp} instances for a 32-bit \gls{isa} and the
respective attack complexities.\label{table:scfp_instances}}
\begin{minipage}{\textwidth}
\begin{center}
 \begin{tabular}{l | c c | c | c c c c | c }
                             & \multicolumn{2}{c|}{Parameters [bit]}  &
Cryptographic &
\multicolumn{4}{c|}{Attack Complexity [bit]} & \\
   Permutation               & $x$ & $s_p$ &
Security [bit] \footnote{Random collisions in the capacity $x$ have
birthday-bound complexity and limit the achievable authenticity,
i.e.,~$x/2$ bits.} &
CIA \footnote{Requires the recovery of capacity and permutation key,
i.e.,~$x + s_p$ bits. \label{fn:state_permutationkey}} &
CRA \footnote{Requires to find and inject the correct patch values, i.e.,~$x$
bits. \label{fn:state_size}} &
ESIP \footref{fn:state_permutationkey}  &
FAIS \footref{fn:state_size} & Type \\\hline
   \mbox{Keccak-$p$[200,12]} & 168 &  --- & 84 & 168  & 168 & 168 & 168 &
\glstext{aee} \\
   \mbox{Keccak-$p$[50,12]}  & 16  &  --- &  8 & 16   & 16  & 16  & 16  &
\glstext{ie} \\
   PRINCE                    & 32  &   96 & 16 & 128  & 32  & 128 & 32  &
\glstext{aeelight} \\
 \end{tabular}
 \footnotetext[4]{CIA: Code-Injection Attacks; CRA: Code-Reuse Attacks;
ESIP: Extraction of Software IP; FAIS: Fault Attack with Instruction Skips}
\end{center}
\end{minipage}
\end{table*}

\subsection{Keyed Permutations}

\gls{aee} enforces its security properties by using a sufficiently large
permutation and thus capacity. However, a sponge capacity providing
cryptographic security levels also implies larger \gls{aee} patch values and
thus memory overhead. On the other hand, \gls{ie} yields lower memory overhead
by using a small permutation, but cannot sufficiently protect software \gls{ip}
and its authenticity. For this reason, an \gls{scfp} instance with low memory
overhead, but with similar security properties as \gls{aee}, is desirable. One
approach to tackle this problem are keyed permutations.

When using a keyed permutation, the security of \gls{scfp} does not
only depend on the sponge capacity $x$, but also on the security level $s_p$ of
the permutation itself. As for \gls{aee} and \gls{ie}, the authenticity when
using keyed permutations is determined by the sponge capacity $x$, i.e.,~the
authenticity level is $x/2$\,bits. However, the complexity of learning the
plaintext of the encrypted binary is $2^{x + s_p}$ and thus also depends on the
security guarantees of the permutation with respect to the permutation key.

\subsubsection{\glsdesc{aeelight}}

We build on this observation and introduce \gls{aeelight} to denote \gls{scfp}
instances based on keyed permutations.
\gls{aeelight} offers the same security bounds as \gls{aee} with respect to
authenticity and \gls{cfi}. For example, control flow hijacking and
fault attacks on control flow have complexity $2^x$, whereas successful
injection of a single instruction has complexity $2^r$. On the other hand,
successful recovery of the software \gls{ip} or the internal state from the
encrypted image has complexity $2^{x+s_p}$. By using a permutation with
sufficiently high security level $s_p$, the confidentiality of software
\gls{ip} is hence guaranteed and state recovery, code injection, and meaningful
forgery are prevented. In particular, even if an attacker recovers the $x$-bit
internal state, meaningful injection or forgery of more than one instruction
still has complexity $2^{s_p}$ as the permutation key is unknown to the
attacker.

For 32-bit instructions, a suitable choice for the keyed permutation is the
64-bit block cipher PRINCE~\cite{DBLP:conf/asiacrypt/BorghoffCGKKKLNPRRTY12},
which uses a 128-bit key to offer $s_p=96$-bit security. This results in a
sponge capacity $x=32$\,bits. State recovery using this \gls{aeelight} instance
has complexity $2^{128}$ and is thus infeasible. This effectively protects the
software \gls{ip} and prevents both code injection and analysis. Contrary to
that, \gls{ie} from before uses a similarly small permutation but cannot
guarantee any of these features without further techniques to hide the
encrypted binary.
However, while the cryptographic level of authenticity guaranteed by this
instance of
\gls{aeelight} is only 16\,bits, meaningful code-reuse attacks and forgery are
much harder. Namely, the expected complexity to find the correct patch value is
$2^{32}$, which is enough to enforce \gls{cfi} and to prevent code-reuse and
physical fault attacks. Besides, the $s_p=96$-bit security of the permutation
further hardens any attempts to tamper with the software binary in a meaningful
way. In particular, even though one single instruction can be manipulated with
complexity $2^{32}$, meaningful modification of multiple instructions is
significantly harder since both the internal \gls{aeelight} state and the
permutation key are unknown to the attacker.

\subsection{Discussion}

\autoref{table:scfp_instances} summarizes the exemplary instances of \gls{aee},
\gls{aeelight}, and \gls{ie}. In detail, \autoref{table:scfp_instances} shows
the respective cryptographic security as well as the actual attack complexities for code injection, code-reuse attacks,
extraction of software \gls{ip}, and instruction skips using fault attacks.
\gls{aee} is the strongest variant with 168-bit security for all considered
attacks. At the further end, \gls{ie} is the smallest variant and offers merely
16-bit security for the mentioned attacks. However, this suffices to enforce
\gls{cfi} and prevent code reuse as well as fault attacks on control flow when
the code binary remains hidden. As a trade-off between these two, \gls{aeelight}
uses keyed permutations to simultaneously attain small 32-bit patch values,
i.e.,~low memory overhead, and good security properties. In particular,
\gls{aeelight} provides 128-bit security in terms of \gls{ip} recovery and code
injection, whereas its security level with respect to code reuse and
control-flow fault attacks is 32\,bits and thus sufficiently high for \gls{cfi}.

\section{Evaluation}
\label{sec:evaluation}

While \gls{scfp} protects software and its execution from a large set of
attacks, \gls{scfp} also has an impact on performance in practice.
In this section, we determine this performance impact by analyzing binary size
increase and execution time overhead of \gls{aeelight} implemented on a
\mbox{RISC-V} processor. Our results demonstrate the practicality of \gls{scfp}
with a size overhead of 19.8\,\% and a performance overhead of 9.1\,\% on
average.

\subsection{Architecture}

The basis of our evaluation hardware is the RI5CY core which is part of the
PULPino~\cite{srcPulpino2017} \gls{soc}. RI5CY is an in-order implementation of
the \mbox{RISC-V} \gls{isa}~\cite{WatermanRISCV2017} with four pipeline stages.
More concretely, our processor supports the RV32IM subset of the \gls{isa} with
privilege architecture version~1.9 and has been extended with optional
\gls{scfp} support. When \gls{scfp} is enabled, \gls{ae} decryption is performed
in an additional pipeline stage between instruction fetching and decoding. The
processor therefore has four pipeline stages when \gls{scfp} is disabled and
five when \gls{scfp} is active.

As \gls{scfp} instance, an \gls{aeelight} configuration with a
64-bit sponge state and the PRINCE block cipher as a keyed permutation in an
APE-like sponge construction has been implemented. Since our processor supports
both normal code execution and \gls{scfp}-protected execution, seven
custom control-flow instructions have been added. These new instructions have
similar semantics as the existing conditional branch (\texttt{BEQ, BNE, BLT,
BLTU, BGE, BGEU}) and jump (\texttt{JAL, JALR}) instructions, but additionally
apply patch values as needed. In particular, our protected conditional branch
instructions (\texttt{BPEQ, BPNE, BPLT, BPLTU, BPGE, BPGEU}) always apply a
patch when the branch is taken. On the other hand, our protected jump
instructions (\texttt{JALP, JALRP}) apply either one or two patch values
depending on the alignment of the target address. The respective patch values
used by these instructions are embedded within the encrypted binary as constants
in the \texttt{.text} section and are located next to the branch/jump
instruction they belong to.

We furthermore designed a \gls{soc}, containing our modified processor with a
design frequency of 100\,MHz, in a UMC 65\,nm technology, which is currently
manufactured as an \gls{asic}. Within this \gls{asic}, our processor occupies
90\,kGE of which around 32\,\% are due to the \gls{scfp} implementation. With
91\,\%, the majority of this overhead is due to the fully unrolled, single cycle
PRINCE implementation. Note that extending the \mbox{RISC-V} core with
\gls{scfp} did not change the target frequency of 100\,MHz.

\subsection{Software Translation}

Our toolchain, which generates binaries for our custom processor in
\gls{aeelight} mode, is at the moment rather simple. We employ the standard
\mbox{RISC-V} GNU toolchain which we only extended with assembling support for our
custom instructions. Additionally, a post-processing tool has been developed
which consumes the final elf binary in order to perform the encryption of the
code and to fill in the required patch values.

Since the C compiler has not been extended with \gls{scfp} support, this
toolchain natively can only handle assembler code which contains placeholders
for the patch values and uses our protected control-flow instructions. However,
due to the way we designed our instruction-set extension, we can quite easily
support the protection of C programs via simple textual replacement of
instructions on the assembly level.

More concretely, at the moment, when compiling C code for our processor, we
first compile the C code to assembly, where we replace all ordinary control-flow
instructions through the protected counterparts and embed \texttt{NOP}
instructions as placeholders for the patch values. The resulting assembly files
are then assembled and linked. Finally, the resulting elf file is processed
using our post-processing tool which emits the encrypted binary.

This simple flow already suffices to demonstrate the practicality of \gls{scfp}.
Note however, given that the compiler is not aware of \gls{scfp}, also the code
has not been optimized for the correct cost model (e.g.,~loops get more
expensive due to the patching). We therefore consider the proper integration of
\gls{scfp} into the compiler as future work and expect that even better
performance can be achieved with an optimized compiler which is aware of
\gls{scfp}.

\subsection{Results}

We used our software toolchain to instrument, compile and encrypt a set of C
benchmarks to evaluate our implementation of \gls{aeelight}. As benchmarks, we
used several programs from the PULPino repository~\cite{srcPulpino2017}:
AES in CBC mode (\texttt{aes\_cbc}), a 2-dimensional matrix convolution
(\texttt{conv2d}), 100 runs of \texttt{dhrystone}, a finite response
filter (\texttt{fir}), a fast Fourier transform (\texttt{fft}), and an
implementation of the inflection point method (\texttt{ipm}). Moreover,
we used two implementation variants of the elliptic curve point
multiplication (SECP192R1) that were internally available at our department.
Both \texttt{ecc} and \texttt{ecc\_opt} are pure C implementations
targeted at microcontrollers. However, while \texttt{ecc} uses a generic
implementation of the underlying multi-precision integer arithmetic, the
multi-precision integer arithmetic in \texttt{ecc\_opt} uses completely
unrolled loops and only works for the specific elliptic curve. We compiled all
programs at optimization level \texttt{-O3}. Since the manufactured \gls{asic}
is not yet available at our department, all runtime values have been determined
using cycle accurate HDL simulation.

\begin{table}[t]
 \caption{Evaluation results of \gls{aeelight} in HDL simulation.
\label{table:aeelight_results}}
 \centering
 \begin{tabular}{l | c c | c c }
                             & \multicolumn{2}{c|}{Code Size (\texttt{text} + \texttt{data})}  &
\multicolumn{2}{c}{Runtime} \\
 & Baseline & Overhead & Baseline & Overhead \\
 & [kB] & [\%] & [kCycles] & [\%] \\ \hline
 \texttt{aes\_cbc}  & 10.0 & 14.8 & 43.4 &  9.5 \\
 \texttt{conv2d}    &  4.6 & 25.6 &  5.4 &  4.8 \\
 \texttt{dhrystone} &  7.5 & 20.1 & 50.6 & 14.4 \\
 \texttt{ecc}       &  9.3 & 21.0 & 4282 &  9.2 \\
 \texttt{ecc\_opt}  &  9.6 & 20.1 & 3032 &  3.8 \\
 \texttt{fir}       &  5.5 & 20.9 & 24.0 &  9.5 \\
 \texttt{fft}       &  7.1 & 16.8 & 45.6 &  7.0 \\
 \texttt{ipm}       &  8.8 & 19.4 &  4.5 & 14.9 \\ \hline
 Average            &      & 19.8 &      &  9.1
 \end{tabular}
\end{table}

\autoref{table:aeelight_results} shows the results of our evaluation of code
size and execution time. In particular, \autoref{table:aeelight_results}
compares the unprotected, standard executables of our benchmark programs with
the executables protected and encrypted via our instance of \gls{aeelight}. Both
program versions have been executed on our modified processor which features
either a four stage (i.e.,~baseline) or a five stage (i.e.,~\gls{aeelight})
pipeline.

For our set of benchmarks, it shows that the overhead in code size due to the
inserted patch values ranges between 14.8\,\% and 25.6\,\% and averages to
19.8\,\%.
This overhead is mainly affected by the number of branches and function calls
in the binary. On the other hand, the runtime overhead ranges between 3.8\,\% and
14.9\,\% and averages to 9.1\,\%. This runtime overhead is significantly lower
than the code size overhead and mainly depends on the number of branches and
function calls that are effectively taken during runtime. This becomes
especially visible for the two implementations of elliptic curves. Namely, as
the inner loops are unrolled in \texttt{ecc\_opt}, the number of executed
branches drops massively from 170\,k taken branches to 20\,k and hence the runtime
overhead for \texttt{ecc\_opt} is much lower than for \texttt{ecc}. On the other
hand, \texttt{ecc} and \texttt{ecc\_opt} yet have very similar code size and
code overhead.

Summarizing, the figures in \autoref{table:aeelight_results} indicate that
\gls{aeelight} can protect against a wide range of \gls{iot} threats with
practical overheads. In addition, our \gls{aeelight} instance features a low
expected
detection latency for execution errors of merely two cycles, because the
\mbox{RISC-V} instruction set we implemented is quite sparse
($\mathtt{\sim}$75\,\% invalid encodings). However, note that using a more
sophisticated toolchain for generating \gls{aeelight} executables can lead to
even smaller overheads.

\section{Related Work}
\label{sec:related_work}

In general, other \gls{cfi} schemes as well as work on \glspl{tee} can be
considered related to \gls{scfp}. However, most techniques provide vastly
different security guarantees given that they are designed to counteract
software attacks only.


\emph{\glsdesc{cfi}.}
Numerous \gls{cfi} schemes have been introduced in the last 30 years. However,
techniques providing fine-grained \gls{cfi} and code integrity, as required
to detect physical attacks, are quite rare~\cite{DBLP:journals/corr/ClercqV17}.

SOFIA~\cite{DBLP:conf/date/ClercqKC0MBPSV16,DBLP:journals/compsec/ClercqGUMV17}
is probably the most relevant technique from this category. SOFIA does not only
encrypt the program, but prevents the execution of any tampered instruction via
MAC checks. Unfortunately, this is rather costly. The average overhead of the
PRINCE-based SOFIA implementation with 64-bit tags in terms of clock cycles and
code size is 141\,\% and 203\,\%, respectively. On the other hand, an
\gls{aeelight} implementation with 64-bit sponge capacity is expected to yield
less overheads, i.e.,~roughly 20\,\% and 40\,\% for runtime and code size,
respectively. Note that, considering that the number and position of patches
does not change when the sponge state/patch size is increased, extrapolating our
results for a given program provides an exact overhead number for the program
size and a plausible estimate for the performance overhead.

Another common approach to enforce \gls{cfi} is to augment the processor
with hardware
monitors~\cite{DBLP:journals/tvlsi/AroraRRJ06,DBLP:journals/tc/MaoW10,
DBLP:conf/cardis/WernerWM15}. Typically, these monitors continuously check that
the processor behaves as expected and raise an alert when an error is observed.
The disadvantage of this approach is that deciding between correct and incorrect
behavior (1-bit of information) effectively introduces a single point of failure
for the error detection. As a result, implementing a reliable monitor becomes a
challenge on its own. Additionally, these techniques can only provide
integrity/authenticity and do not offer confidentiality.

\emph{\glsdesc{tee}.}
\glspl{tee} typically also provide confidentiality and authenticity
for code. However, \glspl{tee} operate on a completely different
level of granularity than \gls{scfp}. The authenticated encryption in
Intel SGX~\cite{mckeen_innovative_2013}, for example, only ensures that the
code and data in memory is protected against tampering. Physical faults in
caches or on processor internal buses, on the other hand, are still possible.
Also, SGX does not prevent common software attack techniques like code-reuse
attacks within enclaves. Therefore, additional \gls{cfi} schemes are needed to
reach similar properties as \gls{scfp} for code. 

\section{Conclusion}
\label{sec:conclusion}

\gls{iot} devices are exposed to a wide range of attacks, such as code
injection, code-reuse attacks, fault attacks, and \gls{ip} theft. While there
are suitable countermeasures for each of these attacks, nowadays' \gls{iot}
devices hardly implement any protection mechanism. On the other hand, it
requires several of the existing countermeasures to mitigate all of the
mentioned attacks. However, a combination of different countermeasures is hard
to analyze and may result in overheads that are too large for \gls{iot} devices.

To overcome this limitation, this work introduced \glsdesc{scfp}
(\glstext{scfp}). \gls{scfp} uses sponge-based authenticated encryption to
encrypt and authenticate software with instruction-level granularity. During
runtime, a hardware extension continuously decrypts instructions at the latest
possible point before the processor's decode stage. As a result, \gls{scfp}
effectively protects confidentiality and authenticity of the software \gls{ip},
and provides fine-grained \gls{cfi} to prevent code injection, code reuse, and
control-flow fault attacks on the processor chip. The \gls{cfi} enforced by
\gls{scfp} is compatible with interrupts and standard operating systems.
To emphasize the flexibility of
\gls{scfp}, we further introduced three different instances of \gls{scfp} for
different application purposes. While \gls{aee} provides all security features
at cryptographic levels of security, \gls{aeelight} reduces the level of
software authenticity in trade for smaller memory overhead. In
addition, \gls{ie} is a very lightweight \gls{cfi} scheme without any guarantees
w.r.t. software authenticity and confidentiality.
Finally, we demonstrated the practicality of \gls{scfp} by extending a
\mbox{RISC-V} processor core with an instance of \gls{aeelight} and evaluating a
set of benchmarks. Our evaluations indicate that \gls{aeelight} is suitable for
many \gls{iot} scenarios with low code size and runtime overheads of 19.8\,\%
and 9.1\,\% on average, respectively.

\section*{Acknowledgements}
We thank Florian Mendel and Christoph Dobraunig for their helpful comments and
many fruitful discussions. Furthermore, we thank Frank K. Gürkaynak, Germain
Haugou, Beat Muheim, and Prof. Luca Benini from the Institute for Integrated
Systems (IIS) at ETH Zurich for supervising David Schaffenrath during his time
at ETH Zurich and for manufacturing the \gls{asic}.

\begin{wrapfigure}{r}{105pt}
  \includegraphics[width=120pt,right]{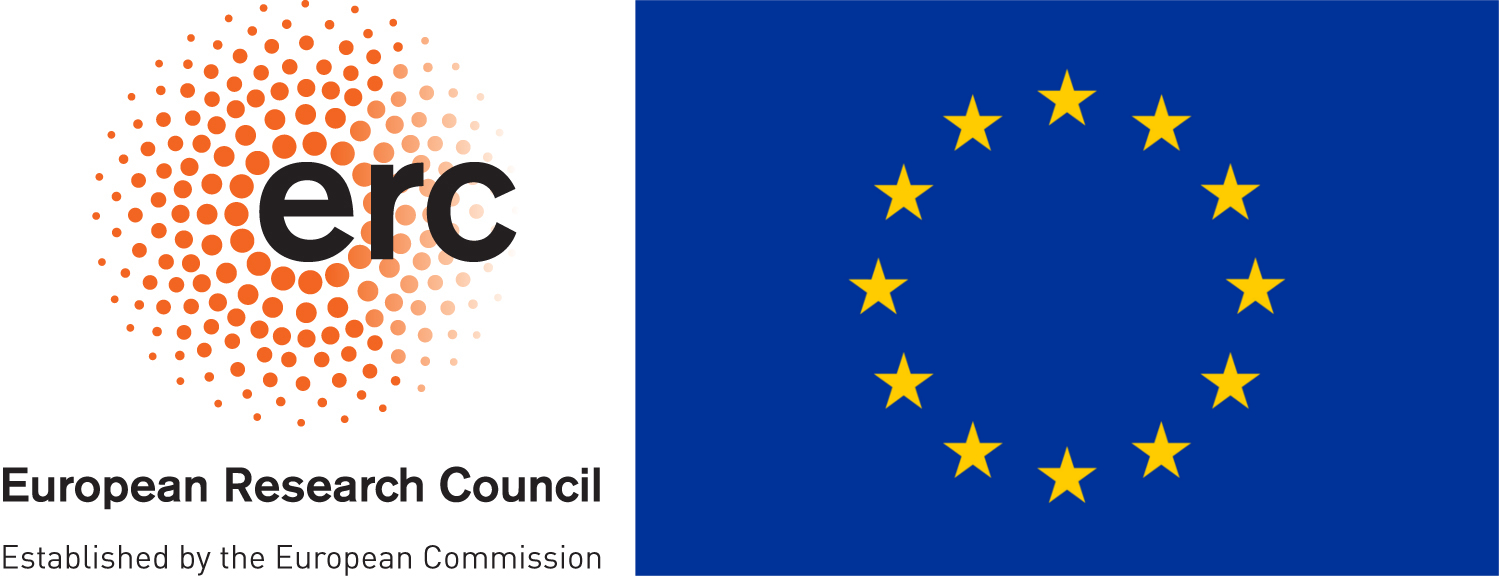}
\end{wrapfigure}

This project has received funding from the European Research Council (ERC) under
the European Union's Horizon 2020 research and innovation programme (grant
agreement No 681402).

\balance

\bibliographystyle{IEEEtranS}
\bibliography{IEEEabrv,main}

\begin{thebibliography}{10}
\providecommand{\url}[1]{#1}
\csname url@samestyle\endcsname
\providecommand{\newblock}{\relax}
\providecommand{\bibinfo}[2]{#2}
\providecommand{\BIBentrySTDinterwordspacing}{\spaceskip=0pt\relax}
\providecommand{\BIBentryALTinterwordstretchfactor}{4}
\providecommand{\BIBentryALTinterwordspacing}{\spaceskip=\fontdimen2\font plus
\BIBentryALTinterwordstretchfactor\fontdimen3\font minus
  \fontdimen4\font\relax}
\providecommand{\BIBforeignlanguage}[2]{{%
\expandafter\ifx\csname l@#1\endcsname\relax
\typeout{** WARNING: IEEEtranS.bst: No hyphenation pattern has been}%
\typeout{** loaded for the language `#1'. Using the pattern for}%
\typeout{** the default language instead.}%
\else
\language=\csname l@#1\endcsname
\fi
#2}}
\providecommand{\BIBdecl}{\relax}
\BIBdecl

\bibitem{MiraiBotnetSource2012}
\BIBentryALTinterwordspacing
``Mirai botnet,'' 2016. [Online]. Available:
  \url{http://github.com/jgamblin/Mirai-Source-Code}
\BIBentrySTDinterwordspacing

\bibitem{abadi_controlflow_2009}
M.~Abadi, M.~Budiu, {\'U}.~Erlingsson, and J.~Ligatti, ``Control-flow
  {{Integrity Principles}}, {{Implementations}}, and {{Applications}},''
  \emph{ACM Trans. Inf. Syst. Secur.}, vol.~13, no.~1, pp. 4:1--4:40, Nov.
  2009.

\bibitem{andreeva_primates_2014}
\BIBentryALTinterwordspacing
E.~Andreeva, B.~Bilgin, A.~Bogdanov, A.~Luykx, F.~Mendel, B.~Mennink, N.~Mouha,
  Q.~Wang, and K.~Yasuda, ``{{PRIMATEs}} v1.02 {{Submission}} to the {{CAESAR
  Competition}},'' Sep. 2014. [Online]. Available: \url{http://primates.ae/}
\BIBentrySTDinterwordspacing

\bibitem{andreeva_ape_2014}
E.~Andreeva, B.~Bilgin, A.~Bogdanov, A.~Luykx, B.~Mennink, N.~Mouha, and
  K.~Yasuda, ``\BIBforeignlanguage{en}{{{APE}}: {{Authenticated
  Permutation-Based Encryption}} for {{Lightweight Cryptography}}},'' in
  \emph{\BIBforeignlanguage{en}{Fast {{Software Encryption}}}}, ser.
  LNCS.\hskip 1em plus 0.5em minus 0.4em\relax {Springer Berlin Heidelberg},
  Mar. 2014, no. 8540, pp. 168--186.

\bibitem{andreeva_security_2015}
E.~Andreeva, J.~Daemen, B.~Mennink, and G.~V. Assche,
  ``\BIBforeignlanguage{English}{Security of {{Keyed Sponge Constructions
  Using}} a {{Modular Proof Approach}}},'' in
  \emph{\BIBforeignlanguage{English}{Fast {{Software Encryption}}}}, ser.
  LNCS.\hskip 1em plus 0.5em minus 0.4em\relax {Springer Berlin Heidelberg},
  2015, vol. 9054, pp. 364--384.

\bibitem{IndustroyerWhitePaper2017}
\BIBentryALTinterwordspacing
{Anton Cherepanov, ESET}, ``Win32/industroyer: A new threat for industrial
  control systems,'' 2017. [Online]. Available:
  \url{https://www.welivesecurity.com/wp-content/uploads/2017/06/Win32_Industroyer.pdf}
\BIBentrySTDinterwordspacing

\bibitem{DBLP:journals/tvlsi/AroraRRJ06}
\BIBentryALTinterwordspacing
D.~Arora, S.~Ravi, A.~Raghunathan, and N.~K. Jha, ``Hardware-assisted run-time
  monitoring for secure program execution on embedded processors,''
  \emph{{IEEE} Trans. {VLSI} Syst.}, vol.~14, no.~12, pp. 1295--1308, 2006.
  [Online]. Available: \url{https://doi.org/10.1109/TVLSI.2006.887799}
\BIBentrySTDinterwordspacing

\bibitem{aumasson_norx_2014}
J.-P. Aumasson, P.~Jovanovic, and S.~Neves, ``{{NORX}} v1,'' Mar. 2014.

\bibitem{DBLP:journals/iacr/Bar-ElCNTW04}
H.~Bar{-}El, H.~Choukri, D.~Naccache, M.~Tunstall, and C.~Whelan, ``The
  sorcerer's apprentice guide to fault attacks,'' 2004, {IACR} Cryptology
  ePrint Archive, Report 2004/100.

\bibitem{bernstein_caesar_2016}
\BIBentryALTinterwordspacing
D.~J. Bernstein, ``{{CAESAR}}: {{Competition}} for {{Authenticated
  Encryption}}: {{Security}}, {{Applicability}}, and {{Robustness}},'' Jan.
  2016. [Online]. Available: \url{http://competitions.cr.yp.to/caesar.html}
\BIBentrySTDinterwordspacing

\bibitem{bertoni_duplexing_2011}
G.~Bertoni, J.~Daemen, M.~Peeters, and G.~{Van Assche}, ``Duplexing the sponge:
  single-pass authenticated encryption and other applications,'' 2011,
  cryptology ePrint Archive, Report 2011/499.

\bibitem{bertoni_security_2011}
\BIBentryALTinterwordspacing
------, ``On the security of the keyed sponge construction,'' \emph{SKEW},
  2011. [Online]. Available: \url{http://sponge.noekeon.org/SpongeKeyed.pdf}
\BIBentrySTDinterwordspacing

\bibitem{bertoni_ketje_2014}
\BIBentryALTinterwordspacing
G.~Bertoni, J.~Daemen, M.~Peeters, G.~{Van Assche}, and R.~{Van Keer},
  ``{{CAESAR}} submission: {{KETJE}} v1,'' Mar. 2014. [Online]. Available:
  \url{http://ketje.noekeon.org/}
\BIBentrySTDinterwordspacing

\bibitem{bertoni_keyak_2015}
\BIBentryALTinterwordspacing
------, ``{{CAESAR}} submission: {{KEYAK}} v2,'' Dec. 2015. [Online].
  Available: \url{http://keyak.noekeon.org/}
\BIBentrySTDinterwordspacing

\bibitem{bletsch_jumporiented_2011}
T.~Bletsch, X.~Jiang, V.~W. Freeh, and Z.~Liang, ``Jump-oriented
  {{Programming}}: {{A New Class}} of {{Code}}-reuse {{Attack}},'' in
  \emph{Proceedings of the 6th {{ACM Symposium}} on {{Information}},
  {{Computer}} and {{Communications Security}}}, ser. ASIACCS '11.\hskip 1em
  plus 0.5em minus 0.4em\relax New York, NY, USA: {ACM}, 2011, pp. 30--40.

\bibitem{DBLP:conf/asiacrypt/BorghoffCGKKKLNPRRTY12}
\BIBentryALTinterwordspacing
J.~Borghoff, A.~Canteaut, T.~G{\"{u}}neysu, E.~B. Kavun, M.~Knezevic, L.~R.
  Knudsen, G.~Leander, V.~Nikov, C.~Paar, C.~Rechberger, P.~Rombouts, S.~S.
  Thomsen, and T.~Yal{\c{c}}in, ``{PRINCE} - {A} low-latency block cipher for
  pervasive computing applications - extended abstract,'' in \emph{Advances in
  Cryptology - {ASIACRYPT} 2012 - 18th International Conference on the Theory
  and Application of Cryptology and Information Security, Beijing, China,
  December 2-6, 2012. Proceedings}, ser. Lecture Notes in Computer Science,
  X.~Wang and K.~Sako, Eds., vol. 7658.\hskip 1em plus 0.5em minus 0.4em\relax
  Springer, 2012, pp. 208--225. [Online]. Available:
  \url{https://doi.org/10.1007/978-3-642-34961-4_14}
\BIBentrySTDinterwordspacing

\bibitem{DBLP:conf/uss/CarliniBPWG15}
N.~Carlini, A.~Barresi, M.~Payer, D.~Wagner, and T.~R. Gross, ``Control-{{Flow
  Bending}}: {{On}} the {{Effectiveness}} of {{Control-Flow Integrity}},'' in
  \emph{24th {{USENIX Security Symposium}} ({{USENIX Security}} 15)}.\hskip 1em
  plus 0.5em minus 0.4em\relax Washington, D.C.: {USENIX Association}, Aug.
  2015, pp. 161--176.

\bibitem{DBLP:conf/sp/ClementsASSKBP17}
\BIBentryALTinterwordspacing
A.~A. Clements, N.~S. Almakhdhub, K.~S. Saab, P.~Srivastava, J.~Koo, S.~Bagchi,
  and M.~Payer, ``Protecting bare-metal embedded systems with privilege
  overlays,'' in \emph{2017 {IEEE} Symposium on Security and Privacy, {SP}
  2017, San Jose, CA, USA, May 22-26, 2017}.\hskip 1em plus 0.5em minus
  0.4em\relax {IEEE} Computer Society, 2017, pp. 289--303. [Online]. Available:
  \url{https://doi.org/10.1109/SP.2017.37}
\BIBentrySTDinterwordspacing

\bibitem{DBLP:conf/uss/Cowan98}
\BIBentryALTinterwordspacing
C.~Cowan, ``Stackguard: Automatic adaptive detection and prevention of
  buffer-overflow attacks,'' in \emph{Proceedings of the 7th {USENIX} Security
  Symposium, San Antonio, TX, USA, January 26-29, 1998}, A.~D. Rubin, Ed.\hskip
  1em plus 0.5em minus 0.4em\relax {USENIX} Association, 1998. [Online].
  Available:
  \url{https://www.usenix.org/conference/7th-usenix-security-symposium/stackguard-automatic-adaptive-detection-and-prevention}
\BIBentrySTDinterwordspacing

\bibitem{DBLP:journals/compsec/ClercqGUMV17}
\BIBentryALTinterwordspacing
R.~de~Clercq, J.~G{\"{o}}tzfried, D.~{\"{U}}bler, P.~Maene, and I.~Verbauwhede,
  ``{SOFIA:} software and control flow integrity architecture,''
  \emph{Computers {\&} Security}, vol.~68, pp. 16--35, 2017. [Online].
  Available: \url{https://doi.org/10.1016/j.cose.2017.03.013}
\BIBentrySTDinterwordspacing

\bibitem{DBLP:conf/date/ClercqKC0MBPSV16}
\BIBentryALTinterwordspacing
R.~de~Clercq, R.~D. Keulenaer, B.~Coppens, B.~Yang, P.~Maene, K.~D. Bosschere,
  B.~Preneel, B.~D. Sutter, and I.~Verbauwhede, ``{SOFIA:} software and control
  flow integrity architecture,'' in \emph{2016 Design, Automation {\&} Test in
  Europe Conference {\&} Exhibition, {DATE} 2016, Dresden, Germany, March
  14-18, 2016}, L.~Fanucci and J.~Teich, Eds.\hskip 1em plus 0.5em minus
  0.4em\relax {IEEE}, 2016, pp. 1172--1177. [Online]. Available:
  \url{http://ieeexplore.ieee.org/document/7459489/}
\BIBentrySTDinterwordspacing

\bibitem{DBLP:journals/corr/ClercqV17}
\BIBentryALTinterwordspacing
R.~de~Clercq and I.~Verbauwhede, ``A survey of hardware-based control flow
  integrity {(CFI)},'' \emph{CoRR}, vol. abs/1706.07257, 2017. [Online].
  Available: \url{http://arxiv.org/abs/1706.07257}
\BIBentrySTDinterwordspacing

\bibitem{dobraunig_ascon_2015}
\BIBentryALTinterwordspacing
C.~Dobraunig, M.~Eichlseder, F.~Mendel, and M.~Schl{\"a}ffer, ``{{ASCON}} v1.1
  {{Submission}} to the {{CAESAR Competition}},'' Aug. 2015. [Online].
  Available: \url{http://ascon.iaik.tugraz.at/}
\BIBentrySTDinterwordspacing

\bibitem{srcPulpino2017}
\BIBentryALTinterwordspacing
{ETH Zurich}, ``Pulpino source repository,'' 2017. [Online]. Available:
  \url{https://github.com/pulp-platform/pulpino}
\BIBentrySTDinterwordspacing

\bibitem{falliere2011w32}
N.~Falliere, L.~O. Murchu, and E.~Chien, ``W32. stuxnet dossier,'' \emph{White
  paper, Symantec Corp., Security Response}, vol.~5, no.~6, 2011.

\bibitem{Francillon:2009:DES:1655077.1655083}
\BIBentryALTinterwordspacing
A.~Francillon, D.~Perito, and C.~Castelluccia, ``Defending embedded systems
  against control flow attacks,'' in \emph{Proceedings of the First ACM
  Workshop on Secure Execution of Untrusted Code}, ser. SecuCode '09.\hskip 1em
  plus 0.5em minus 0.4em\relax New York, NY, USA: ACM, 2009, pp. 19--26.
  [Online]. Available: \url{http://doi.acm.org/10.1145/1655077.1655083}
\BIBentrySTDinterwordspacing

\bibitem{free60}
\BIBentryALTinterwordspacing
Free60.org, 2012. [Online]. Available:
  \url{http://free60.org/wiki/Reset_Glitch_Hack}
\BIBentrySTDinterwordspacing

\bibitem{gligoroski_cipher_2015}
D.~Gligoroski, H.~Mihajloska, S.~Samardjiska, H.~Jacobsen, M.~El-Hadedy, R.~E.
  Jensen, and D.~Otte, ``\ensuremath{\pi}\textendash{}{{Cipher}} v2.0,'' Aug.
  2015.

\bibitem{gruss_rowhammerjs_2015}
D.~Gruss, C.~Maurice, and S.~Mangard, ``Rowhammer.js: {{A Remote
  Software-Induced Fault Attack}} in {{JavaScript}},'' \emph{arXiv:1507.06955
  [cs]}, Jul. 2015.

\bibitem{kim_flipping_2014}
Y.~Kim, R.~Daly, J.~Kim, C.~Fallin, J.~H. Lee, D.~Lee, C.~Wilkerson, K.~Lai,
  and O.~Mutlu, ``Flipping bits in memory without accessing them: {{An}}
  experimental study of {{DRAM}} disturbance errors,'' in \emph{2014
  {{ACM}}/{{IEEE}} 41st {{International Symposium}} on {{Computer
  Architecture}} ({{ISCA}})}, Jun. 2014, pp. 361--372.

\bibitem{korak_effects_2014}
T.~Korak and M.~H{\"o}fler, ``On the {{Effects}} of {{Clock}} and {{Power
  Supply Tampering}} on {{Two Microcontroller Platforms}},'' in \emph{2014
  {{Workshop}} on {{Fault Diagnosis}} and {{Tolerance}} in {{Cryptography}}
  ({{FDTC}})}, Sep. 2014, pp. 8--17.

\bibitem{DBLP:conf/osdi/KuznetsovSPCSS14}
\BIBentryALTinterwordspacing
V.~Kuznetsov, L.~Szekeres, M.~Payer, G.~Candea, R.~Sekar, and D.~Song,
  ``Code-pointer integrity,'' in \emph{11th {USENIX} Symposium on Operating
  Systems Design and Implementation, {OSDI} '14, Broomfield, CO, USA, October
  6-8, 2014.}, J.~Flinn and H.~Levy, Eds.\hskip 1em plus 0.5em minus
  0.4em\relax {USENIX} Association, 2014, pp. 147--163. [Online]. Available:
  \url{https://www.usenix.org/conference/osdi14/technical-sessions/presentation/kuznetsov}
\BIBentrySTDinterwordspacing

\bibitem{DBLP:journals/tc/MaoW10}
\BIBentryALTinterwordspacing
S.~Mao and T.~Wolf, ``Hardware support for secure processing in embedded
  systems,'' \emph{{IEEE} Trans. Computers}, vol.~59, no.~6, pp. 847--854,
  2010. [Online]. Available: \url{https://doi.org/10.1109/TC.2010.32}
\BIBentrySTDinterwordspacing

\bibitem{mckeen_innovative_2013}
\BIBentryALTinterwordspacing
F.~McKeen, I.~Alexandrovich, A.~Berenzon, C.~V. Rozas, H.~Shafi, V.~Shanbhogue,
  and U.~R. Savagaonkar, ``Innovative instructions and software model for
  isolated execution,'' in \emph{Proceedings of the 2nd {{International
  Workshop}} on {{Hardware}} and {{Architectural Support}} for {{Security}} and
  {{Privacy}}}.\hskip 1em plus 0.5em minus 0.4em\relax {ACM}, 2013, pp. 1--1.
  [Online]. Available:
  \url{https://software.intel.com/en-us/articles/innovative-instructions-and-software-model-for-isolated-execution}
\BIBentrySTDinterwordspacing

\bibitem{mennink_security_2015}
B.~Mennink, R.~Reyhanitabar, and D.~Viz{\'a}r, ``Security of {{Full-State
  Keyed}} and {{Duplex Sponge}}: {{Applications}} to {{Authenticated
  Encryption}},'' 2015, cryptology ePrint Archive, Report 2015/541.

\bibitem{morawiecki_icepole_2014}
P.~Morawiecki, K.~Gaj, E.~Homsirikamol, K.~Matusiewicz, J.~Pieprzyk,
  M.~Rogawski, M.~Srebrny, and M.~Wojcik, ``{{ICEPOLE}} v1,'' Mar. 2014.

\bibitem{saarinen_stribobr2_2015}
\BIBentryALTinterwordspacing
M.-J.~O. Saarinen and B.~B. Brumley, ``{{STRIBOBr2}}: "{{WHIRLBOB}}" {{Second
  Round CAESAR Algorithm Tweak Specification}},'' Sep. 2015. [Online].
  Available: \url{http://www.stribob.com/}
\BIBentrySTDinterwordspacing

\bibitem{sasaki_how_2015}
Y.~Sasaki and K.~Yasuda, ``How to {{Incorporate Associated Data}} in
  {{Sponge-Based Authenticated Encryption}},'' in \emph{Topics in
  {{Cryptology}} - {{CT-RSA}} 2015}, ser. LNCS, vol. 9048.\hskip 1em plus 0.5em
  minus 0.4em\relax San Francisco, CA, USA: {Springer International
  Publishing}, 2015, pp. 353--370.

\bibitem{shacham_geometry_2007}
H.~Shacham, ``The {{Geometry}} of {{Innocent Flesh}} on the {{Bone}}:
  {{Return}}-into-libc {{Without Function Calls}} (on the x86),'' in
  \emph{Proceedings of the 14th {{ACM Conference}} on {{Computer}} and
  {{Communications Security}}}, ser. CCS '07.\hskip 1em plus 0.5em minus
  0.4em\relax New York, NY, USA: {ACM}, 2007, pp. 552--561.

\bibitem{DBLP:conf/ccs/ShachamPPGMB04}
\BIBentryALTinterwordspacing
H.~Shacham, M.~Page, B.~Pfaff, E.~Goh, N.~Modadugu, and D.~Boneh, ``On the
  effectiveness of address-space randomization,'' in \emph{Proceedings of the
  11th {ACM} Conference on Computer and Communications Security, {CCS} 2004,
  Washington, DC, USA, October 25-29, 2004}, V.~Atluri, B.~Pfitzmann, and P.~D.
  McDaniel, Eds.\hskip 1em plus 0.5em minus 0.4em\relax {ACM}, 2004, pp.
  298--307. [Online]. Available:
  \url{http://doi.acm.org/10.1145/1030083.1030124}
\BIBentrySTDinterwordspacing

\bibitem{team2003pax}
P.~Team, ``Pax address space layout randomization (aslr),'' 2003.

\bibitem{WatermanRISCV2017}
\BIBentryALTinterwordspacing
A.~Waterman and K.~Asanovic, ``{The RISC-V Instruction Set Manual, Volume I:
  User-Level ISA, Version 2.2},'' EECS Department, University of California,
  Berkeley, Tech. Rep., May 2017. [Online]. Available:
  \url{https://content.riscv.org/wp-content/uploads/2017/05/riscv-spec-v2.2.pdf}
\BIBentrySTDinterwordspacing

\bibitem{DBLP:conf/cardis/WernerWM15}
\BIBentryALTinterwordspacing
M.~Werner, E.~Wenger, and S.~Mangard, ``Protecting the control flow of embedded
  processors against fault attacks,'' in \emph{Smart Card Research and Advanced
  Applications - 14th International Conference, {CARDIS} 2015, Bochum, Germany,
  November 4-6, 2015. Revised Selected Papers}, ser. Lecture Notes in Computer
  Science, N.~Homma and M.~Medwed, Eds., vol. 9514.\hskip 1em plus 0.5em minus
  0.4em\relax Springer, 2015, pp. 161--176. [Online]. Available:
  \url{https://doi.org/10.1007/978-3-319-31271-2_10}
\BIBentrySTDinterwordspacing

\bibitem{DBLP:journals/tcad/WilkenS90}
\BIBentryALTinterwordspacing
K.~D. Wilken and J.~P. Shen, ``Continuous signature monitoring: low-cost
  concurrent detection of processor control errors,'' \emph{{IEEE} Trans. on
  {CAD} of Integrated Circuits and Systems}, vol.~9, no.~6, pp. 629--641, 1990.
  [Online]. Available: \url{http://dx.doi.org/10.1109/43.55193}
\BIBentrySTDinterwordspacing

\end{thebibliography}

\end{document}